# Modeling Incoherent Strain Mediated Multiferroic Bennett Clocking

Jin-Zhao Hu,[1] John P. Domann,[2] Qianchang Wang,[1] Cheng-Yen Liang,[1] Scott Keller,[1] Gregory P. Carman,[1] Abdon E. Sepulveda[1]

[1]Department of Mechanical and Aerospace Engineering, University of California, Los Angeles, California, 90095, USA
[2]Department of Biomedical Engineering and Mechanics, Virginia Tech, Blacksburg, Virginia, 24061, USA

Email: abdon.sepulveda@gmail.com

**Abstract**
Strain mediated Bennett clocking has only recently been experimentally demonstrated and suffered from high error rates. Most models used to explain this behavior are macrospin models. Predictions of these models do not match experimental designs since they consider all spins rotating coherently and no magnetoelastic strain feedback. In this paper a fully coupled nonlinear model (LLG plus elastodynamics) was used to simulate voltage induced Bennett clocking. This modelling captures the full spin dynamics as well as shape anisotropy. Two materials were studied (Ni and Terfenol-D) which have very different exchange lengths. The simulation results show that incoherent rotation may occur due to the uniaxial nature of the magnetoelastic coupling.
Keywords: Bennett clocking, meltiferroics, magnetoelectrics, nanomagnetics

**1. Introduction**
Consumer electronics are responsible for approximately 18% of all US home energy consumption, a number predicted to increase in upcoming years.[1] With the increasing proliferation of always-on electronic devices like cell phones and computers, the energy efficiency of traditional CMOS (Complementary Metal Oxide Semiconductor) devices needs to be scrutinized and improved. In the early 2000's CMOS efficiency began rapidly decreasing due to leakage currents in transistors as the devices became smaller. Standby power is now comparable to the actual dynamic operating power in transistors.[2] While memory elements are commonly designed with 40kbT (~0.16 aJ) energy barriers to protect against thermal fluctuations, discharging a single transistor (i.e., flipping a single bit) dissipates around 450aJ.[3] Therefore, nearly 3000 times more energy is dissipated to flip a CMOS bit of information than the energy barrier actually requires (0.03% efficient). Nanomagnetic logic (NML) is proposed as a route to more energy efficient, non-volatile, memory and logic devices. These devices consume drastically less power to flip a bit of information and dissipate zero standby power.

NML evolved from Bennett's early work using quantum cellular automata (QCA) to store and process information.[4] Magnetic quantum cellular automata (MQCA) use magnetization as the information carrier, in contrast to electrostatic QCA devices that rely on charge transport. MQCA has been demonstrated as a viable room temperature technology, while electrostatic QCA devices tend to require cryogenic temperatures for operation.[5] NML is the modern term used to describe the use of MQCA systems,[6,7] and it is the topic of this paper. Figure 1 shows a typical manifestation of NML using ellipsoidal single domain nanomagnets that are coupled via dipolar interactions (as described in Bennett's original work[4]). After the input bit is oriented in the desired position, a clocking field rotates the remaining magnets into a quasi-stable orientation. Removing the clocking field allows dipolar coupling to antiferromagnetically align the magnets. This causes information from the input bit to cascade along the nanomagnet wire. Therefore, controlling the orientation of the input magnet (0 or 1) allows the control an output magnet and transfer data when the system is clocked. This coupling scheme provides the basic mechanism used to transmit and store information in Bennet clocked NML devices.

A key advantage of NML is its low energy consumption. In NML systems, power dissipation occurs in two primary areas: within the magnetic element, and while generating the clocking field. Importantly, single domain magnets change state through near uniform spin rotation, so the energy dissipated within each magnetic element is very small and due primarily to Gilbert damping instead of domain wall motion.[8] In the limit of adiabatic switching, internal losses can become extremely small and have been predicted to approach Landauer's limit (i.e., 3 zJ at room temperature).[9] Hong et al. recently experimentally confirmed these predictions for an array of nanomagnets using a sensitive magnetometry setup.[10] As energy dissipation within the magnet is quite small, Bennett clocking losses are mainly attributable to the generation of the clocking field itself.

The majority of previous NML work has focused on clocking fields created by passing an electric current through a wire. Electric currents have been used to create magnetic fields that clock either all magnets at once,[5,7,10–13] or operate in specific clocking zones.[15] It has been noted that this clocking scheme dissipates orders of magnitude more energy than losses due to Gilbert damping.[12] This is why optimistic estimates only show 10x efficiency improvements over CMOS for this clocking technique.[16] Additionally, this approach is prone to large error rates, as thermal fluctuations and slight manufacturing defects can cause premature bit flips in the middle of long nanomagnet chains.[12,14] While introducing biaxial anisotropy[17] and using concave shapes[18] have both reduced error rates, bit errors still increase as the length of a clocking zone increases and the location of the error is very sensitive to manufacturing defects.[14] This indicates that local / individual control of the nanomagnets is highly desirable, to reduce the clocking zone size and therefore reduce error rates.

Recent efforts have focused on creating energy efficient localized clocking schemes using spin orbit torque, and strain mediated multiferroic heterostructures. A clocking scheme using spin orbit torque was experimentally demonstrated by Bhowmik et al in 2014 that used approximately 100x smaller currents than required for magnetic field based clocking.[19] This corresponds to a 3-4 order of magnitude improvement in energy efficiency. Furthermore, this approach had a 100% success rate, albeit with a very small sample size of only five attempts. Simultaneously, substantial work on strain mediated clocking has indicated it's highly energy efficient nature,[20–26] with recent predictions indicating only 1 aJ may be required to flip a bit of information.[26] Therefore, strain mediated clocking is expected to be 2-3 orders of magnitude more efficient than transistor based logic and 1 order of magnitude better than spin orbit clocking. The rest of this paper will focus on strain mediated clocking.

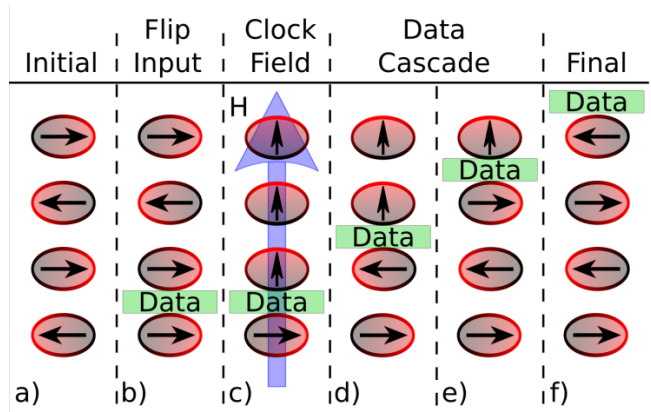

**Figure 1.** Bennett Clocking propagates information along a nanomagnet wire. (a) Initial state from previous computation. (b) The input magnet (bottom magnet), is flipped to a desired value, then (c) a clocking field is applied that causes (d-f) a data cascade due to dipolar interactions, creating (f) a new equilibrium state

Strain mediated Bennett clocking provides highly localized and energy efficient control of NML devices. This approach deposits magnetoelastic nanomagnetic elements onto a piezoelectric substrate. Applying a voltage to nearby electrodes deformation is induced in the substrate which in turn locally strains the nanomagnet and clocks the magnetization with an effective magnetoelastic field. This approach has the advantage of being able to clock single magnets at a time in a highly energy efficient manner. However, strain mediated Bennett clocking has only recently been experimentally demonstrated, and suffered from high error rates.[26] Several models have been created to analyze error rates in strain mediated approaches, but they predominantly rely on macrospin models using the LLG equations with[25] or without[20–24] a stochastic thermal fluctuation field. While the use of macrospin models can be very useful for the initial evaluation of magnetic phenomena, more detailed models are required to aid in the fabrication of practical devices. Macrospin models assume that the magnetization, dipolar coupling field, and applied strain are all uniform throughout each magnetic element, which is a condition not met in actual devices. The predictions of macrospin models can vary substantially from experiments, and more detailed finite element models.[27] This is particularly true for thin film heterostructures, where effects like shear lag lead to nonuniform strain and magnetization profiles, in contradiction with assumptions of uniformity. This paper presents a numerical model that fully couples elastodynamics and micromagnetics to provide an in-depth analysis of strain mediated Bennett clocking. Room temperature is considered with no thermal fluctuations in the model. The model shows how incoherent rotations may result from the uniaxial nature of the magnetoelastic coupling.

## 2. Theory

In this section, a 3D simulation model fully coupling micromagnetics and elastodynamics is described. A synopsis of the system of coupled partial differential equations and numerical methods used to simulate a wide range of geometries are illustrated. For a more detailed derivation the reader is referred to work by Liang et al.[27,28]

The model couples the precessional magnetization dynamics of the Landau-Lifshitz-Gilbert (LLG) equation with mechanical stress and strain governed by elastodynamics. Additionally, the response of the piezoelectric thin film is modeled using linear constitutive equations that relate strain and electric field. The model assumes small elastic deformations (linear elasticity), electrostatics, and negligible contribution from electric currents.

The coupled governing equations used in this work are as follows. The elastodynamics equation governing mechanical stress and displacements is

$$\rho \frac{d\underline{u}^2}{dt^2} = \nabla \cdot \underline{\underline{\sigma}} \qquad \text{(Eq. 1)}$$

where $\rho$ is the mass density, $\underline{\sigma}$ is the Cauchy stress tensor, $\underline{u}$ is the displacement vector, and $t$ is time. Magnetization dynamics are governed by the phenomenological LLG equation,

$$\frac{\partial \underline{m}}{\partial t} = -\mu_0 \gamma \left( \underline{m} \times \underline{H}_{eff} \right) + \alpha \left( \underline{m} \times \frac{\partial \underline{m}}{\partial t} \right) \quad \text{(Eq. 2)}$$

where $\mu_0$ is the permeability of free space, $\gamma$ is the Gilbert gyromagnetic ratio, $\alpha$ is the Gilbert damping constant, and $\underline{m}$ is the normalized magnetization vector. The effective magnetic field, $\underline{H}_{eff}$, includes the externally applied magnetic field ($\underline{H}_{ext}$), exchange field ($\underline{H}_{ex}$), demagnetization field ($\underline{H}_d$), and magnetoelastic field ($\underline{H}_{me}$). Detailed expressions for these terms can be found in the literature.[27,28] The demagnetization field is calculated by using the quasi-static Ampere's law. This leads to $\underline{H}_d = -\nabla \psi$ where $\psi$ is the scalar magnetic potential. Combining this equation with the required $\nabla \cdot \underline{B} = 0$ and the constitutive relation $\underline{B} = \mu_0(\underline{H} + \underline{M})$ produces the governing equation for the magnetic scalar potential in terms of the normalized magnetization.

$$\nabla^2 \psi = M_s (\nabla \cdot \underline{m}) \quad \text{(Eq. 3)}$$

where $M_s$ is the saturation magnetization.

In a similar fashion to the magnetic scalar potential, the electrostatic Faraday's Law implies that $\underline{E} = -\nabla \phi$, where $\phi$ is the electric potential. This equation coupled with Gauss's Law and the linear piezoelectric constitutive equations provides the piezoelectric coupling within the model. Substituting the piezoelectric constitutive relations into the elastodynamics equation (Eq. 1) and LLG equation (Eq. 2) produces a set of cross-coupled non-linear equations containing displacement, magnetization, magnetic scalar potential, and electrical field as shown in (Eq. 4) and (Eq. 5).

$$\rho \frac{d\underline{u}^2}{dt^2} - \nabla \cdot \underline{\underline{C}} \left[ \frac{1}{2} \left( \nabla \underline{u} + (\nabla \underline{u})^T \right) \right] + \nabla \cdot \underline{\underline{C}} \left[ \underline{\underline{\lambda}}^m \underline{m} \underline{m}^T \right] + \nabla \cdot \underline{\underline{C}} \left( \underline{\underline{d}} \underline{E} \right) = 0 \quad \text{(Eq. 4)}$$

$$\frac{\partial \underline{m}}{\partial t} = -\mu_0 \gamma \left( \underline{m} \times \left( \underline{H}_{ext} + \underline{H}_{ex}(\underline{m}) + \underline{H}_d(\psi) + \underline{H}_{me}(\underline{m}, \underline{u}(\underline{E})) \right) \right) + \alpha \left( \underline{m} \times \frac{\partial \underline{m}}{\partial t} \right) \quad \text{(Eq. 5)}$$

where $\underline{\underline{C}}$ is the stiffness tensor, $\underline{\underline{\lambda}}^m$ is the magnetostriction tensor, $\underline{E}$ is the electric field vector, and $\underline{\underline{d}}$ is the piezoelectric coupling tensor. These coupled systems of partial differential equations are solved simultaneously for the mechanical displacement ($u, v, w$), electric potential ($\phi$), magnetic potential ($\psi$), and magnetization ($m_x, m_y, m_z$).

Numerical solution of the magneto-electro-mechanical coupled equations is obtained by using a finite element formulation (implemented in COMSOL) with an implicit backward differentiation (BDF) time stepping scheme. In order to decrease solution time, the system of equations is solved simultaneously using a segregated method, which splits the solution process into sub-steps using a damped Newton's method. This coupled model provides dynamic results for the full strain and micromagnetic spin distribution in the magnetoelastic component coupled with a piezoelectric layer. Convergence studies (i.e., mesh size and time steps) were evaluated to ensure the accuracy of all models.

## 3. Simulation Setup

The objective of the simulation is to analyze the single-domain rotation behavior of Nickel and Terfenol-D nano-ellipses while Bennett clocking. Figure 2 shows the Bennett clocking geometry used in this paper. Four pairs of nano-ellipses and electrodes are perfectly bonded to a 500 nm thick PZT-5H thin film with platinum ground electrode, mounted on a rigid Si/SiO2 substrate. Figure 2a shows a perspective view of the four ellipse/electrode pairs used to form a Bennett clocking wire. Figure 2b shows the in-plane dimensions for each ellipse, which are 100 nm × 90 nm with a 10 nm thickness. Each nano-ellipse is centered between the two 10 nm thick 90 nm square Au electrodes. The center-to-center distance between the electrode and the ellipse is 280 nm. The center-to-center distance between adjacent ellipses is 140 nm (i.e., a gap of 50 nm). Above the PZT layer is 1 μm x 1 μm x 500nm of air to capture field propagation. The Si/SIO2 substrate is assumed mechanically rigid and modeled using fixed boundary conditions in the numerical model. Therefore, the entire model, containing the ellipses, electrodes, piezoelectric substrate, and surrounding air, is 1 μm x 1 μm x 1 μm. Under these dimensions and distance between the elements, the energy required for per flip in the Bennett clocking system is 8.7 fJ ($E = \frac{1}{2}QV$ where $V$ is the voltage on the electrodes and $Q$ is the total charge for the period of time that the voltage is applied on the electrodes).

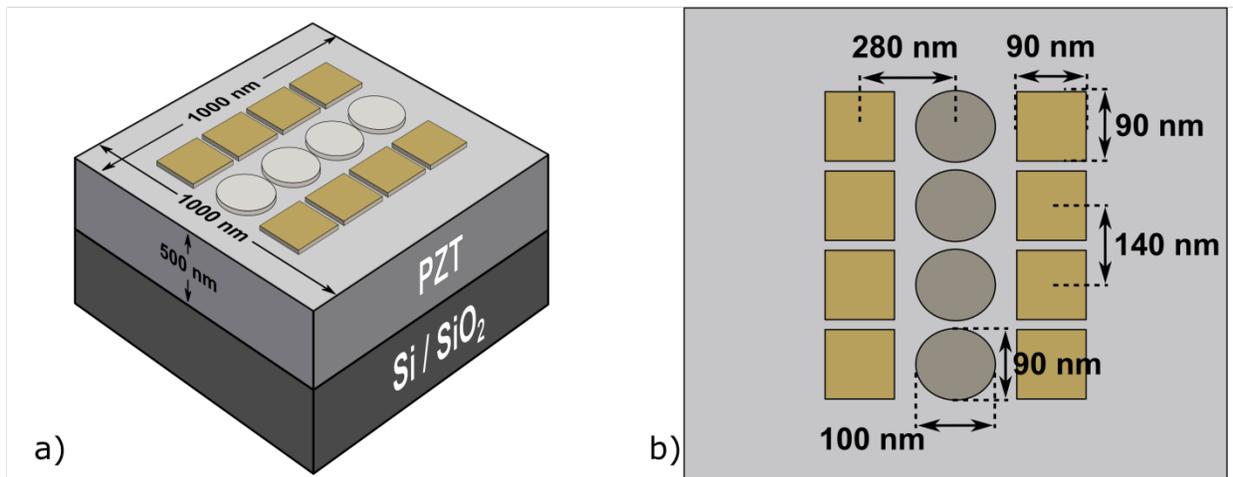

**Figure 2.** The geometry settings of the model. Four ellipses form a line with their electrodes, with the distance between every 2 ellipses is 140 nanometers. Every ellipse has the same aspect ratio (0.9). The 1000nm cross 1000nm's PZT layer lies on a silicon layer. When positive or negative voltage is applied to the electrodes one by one, the PZT layer will produce enough strain to help change the shape anisotropy and drive the circuit.

Two magnetic materials were studied in this paper, the material properties for Nickel and Terfenol-D are shown in Table 1. The Nickel and Terfenol-D nano-ellipses were assumed polycrystalline, therefore crystalline anisotropy is neglected. Our model shows that in this caes, the Gilbert damping do not influence the final states and general trends but dramatically improve the calculation complexity. In this situation, the Gilbert damping ratio is set as $\alpha = 0.5$ to improve model stability and reduce run time for numerical purposes.[29] The final equilibrium state is not affected by this compromise. Transversely isotropic PZT-5H is modeled with piezoelectric coefficient $d_{33} = 5.93 \times 10^{-10}$ C/N, $d_{31} = -2.74 \times 10^{-10}$ C/N, stiffness $c_{11} = c_{22} = 127$ GPa, $c_{12} = 80.2$ GPa, $c_{13} = c_{23} = 84.6$ GPa, $c_{33} = 117$ GPa, $c_{44} = c_{55} = 22.9$ GPa, and density $\rho = 7500$ kg/m³. The z-direction is treated as the transverse c-axis. Young's modulus and Poisson's ratio for the Au electrodes are $E = 70$ GPa and $\nu = 0.44$, respectively.

Table 1. Simulated Material Properties

| Parameter | Description | Units | Nickel | Terfenol-D |
|---|---|---|---|---|
| $M_s$ | Saturation Magnetization | A/m | $4.8 \times 10^5$ | $8 \times 10^5$ |
| $A_{ex}$ | Exchange Stiffness | J/m | $1.05 \times 10^{-11}$ | $1 \times 10^{-11}$ |
| $L_{ex}$ | Exchange Length | nm | 8.5 | 5 |
| $\lambda_s$ | Saturation Magnetostriction | - | $-34 \times 10^{-6}$ | $1200 \times 10^{-6}$ |
| $E$ | Young's Modulus | GPa | 180 | 80 |
| $\rho$ | Density | kg/m³ | 8900 | 9210 |
| $\nu$ | Poisson's Ratio | - | 0.3 | 0.3 |

The boundary conditions and element discretization are as follows. Zero-displacement boundary conditions were applied to all four sides and the bottom surface of the piezoelectric film. The bottom surface was also electrically grounded. The top surface of the piezoelectric layer is free to deform and interact with the Bennett clocking wire. The nano-ellipses are discretized using tetrahedral elements with element size on the order of exchange length $L_{ex} = \sqrt{2A_{ex}/\mu_0 M_s^2}$. The remainder of the structure (i.e., PZT-5H thin film, Au electrodes) is discretized using tetrahedral elements with graded element sizes dependent upon local geometry.

Figure 3 shows a schematic of the strain mediated Bennett Clocking control sequence modeled in this paper. In a classical logic circuit, the goal is to process information from one element to another. Therefore, if ellipse 1 is made to switch from 0 to 1 (i.e., an input signal switches its magnetization from left to right), the goal is to make ellipse 4 switch accordingly. Figure 3a shows the system's initial antiferromagnetic equilibrium state with zero strain (voltage) applied. When ellipse 1 rotates to the right, ellipse 2 does not spontaneously flip to the left because dipole coupling with ellipse 1 is too weak to overcome the combined energy barrier from shape anisotropy and dipole coupling with ellipse 3. To flip ellipse 2, magnetoelastic anisotropy is used to eliminate the energy barrier, and allow dipolar coupling to rotate the magnetization. Figure 3c shows anisotropic strain is applied to ellipse 2 and 3 by applying a voltage to their Au electrodes. This creates an easy axis in the up/down direction, causing the ellipses to rotate as shown in Figure 3c. The same voltage is then applied to ellipse 4 (Figure 3d), while the voltage on ellipse 2 is released. This allows ellipse 2 to antiferromagnetically couple with ellipse 1 and propagates the data forward one ellipse. This same procedure is repeated to clock subsequent ellipses. Figure 3f shows the final equilibrium position following Bennett clocking, illustrating the transfer of information from ellipse 1 to ellipse 4. The simulated clock uses 1V that is sequentially applied to the Au electrodes while the PZT bottom remains electrically grounded (i.e., a 2 MV/m electric field is applied through the PZT film thickness). As Terfenol-D is positive magnetoelastic, and Nickel is negative magnetoelastic, the polarity of the applied voltage is reversed for the two materials.

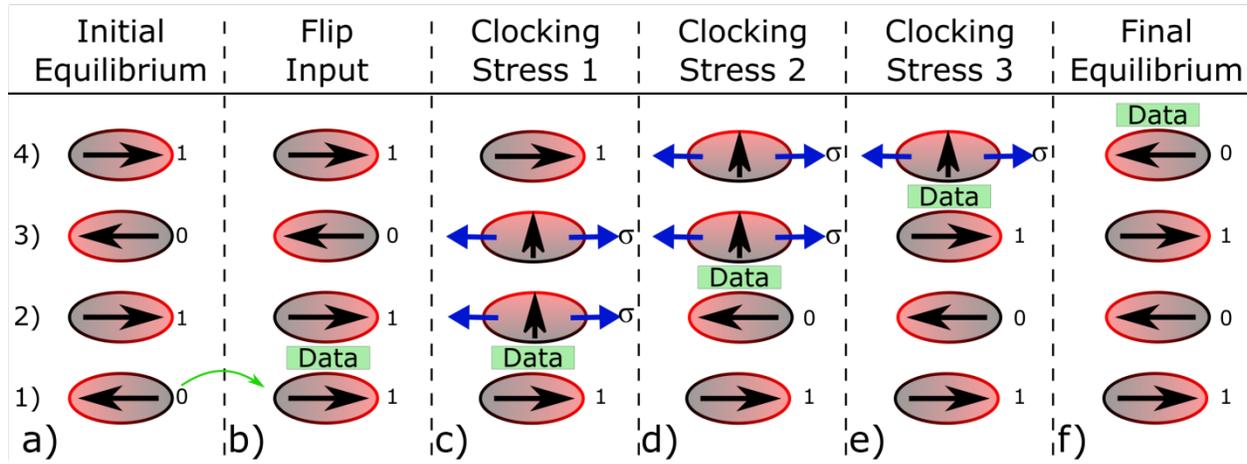

**Figure 3.** Strain mediated Bennett clocking. a) Initial equilibrium position from previous computation. b) An input signal flips ellipse 1 and initiates the clocking sequence. c) Stress is applied to ellipses 2 and 3 to create an easy axis perpendicular to the initial magnetization. d) The voltage is advanced one ellipse, allowing dipolar coupling to flip ellipse 2. e) The process is repeated to flip ellipse 3, and then ellipse 4. f) The final equilibrium positions.

## 4. Results and Discussion

This section presents the dynamic response of the Nickel and Terfenol-D Bennett clocking wires. First the Nickel ellipses will be analyzed, and shown to rotate coherently when strained. Subsequently, the stronger magnetoelastic coupling in Terfenol-D will lead to faster switching times but becomes nonuniformly magnetized and rotates incoherently.

Figure 4a shows the entire Nickel Bennett clocking wire at t=2.5 ns. The magnetic orientation is indicated with red arrows, while the surface color shows the electric potential. When voltage is applied to the relevant electrode pairs, a tensile strain is generated in the x-direction, creating a magnetoelastic easy axis in the y-direction. As a result, the magnetization in ellipses 2 and 3 begins rotating into the y-direction. Based on the voltage distribution shown in the figure, it can clearly be seen that the piezoelectric strain is highly localized on the space between each electrode pair, and does not affect the magnetic orientation of adjacent ellipses.

Figures 4b to 4e show a summary of the entire Bennett clocking sequence for the Nickel ellipses. Figure 4b shows the average rotation angle ($\theta$) for each of the four ellipses as a function of time. Figure 4c shows the voltage on each electrode pair as a function of time. Each ellipse shows a three-stage switching behavior, corresponding to turning the voltage on, the constant voltage dwell time, and turning the voltage off. By t=7 ns, each ellipse has successfully rotated 180 degrees, and the Bennett clocking wire finishes in a stable antiferromagnetic arrangement. It takes approximately 1 ns apiece for ellipses 2 and 3 to rotate 90 degrees when voltage is applied and creates a magnetoelastic easy axis. Another 1 ns is taken to rotate the final 90 degrees when the voltage is turned off, and the combination of dipole coupling, and shape anisotropy forces each ellipse to antiferromagnetically couple. It takes approximately 1 ns for ellipse 4 to rotate when voltage is turned on, but 2 ns to rotate its final 90 degrees. Due to use of a large Gilbert damping coefficient, it is anticipated that actual materials may flip faster than reported here, but also exhibit transient oscillations prior to stabilizing.

Figures 4d and 4e highlight the coherent magnetization rotation in the Nickel ellipses. Figure 4d, shows the magnitude of the maximum average in-plane magnetization component in each ellipse, defined by

$$|m_{avg}| = \sqrt{m_{avg,x}^2 + m_{avg,y}^2 + m_{avg,z}^2} \qquad (Eq.\ 6)$$

where $m_{avg,x}$, $m_{avg,y}$ and $m_{avg,z}$ are the average components in the x, y and z directions. For a coherent / uniform configuration, a magnitude of $|m_{avg}| = 1$ indicates all spins are in-plane and pointing in the same direction. While for an incoherent / nonuniform configuration, spins partially cancel each other and therefore the average component in each direction is reduced. As a result, a reduction $|m_{avg}| < 1$ would indicate that some elements' spins are pointing in different directions. The extreme case is $|m_{avg}| = 0$, which means $|m_{avg,x}| = |m_{avg,y}| = |m_{avg,z}|0$ and all the spins are in random state.[29] It is evident in Figure 4d that the average moment always has a magnitude of 1 for each ellipse, indicating coherent rotation at all time steps. This finding is confirmed in Figure 4e, which highlights ellipse 3 at t = 2.5, 4.5 and 6.5 ns. Each picture indicates the magnetic moments coherently rotate for the Nickel ellipses. The results in Figure 4 indicate a new antiferromagnetic ground state was reached via coherent magnetization rotation, and information was successfully transmitted from ellipse 1 to ellipse 4.

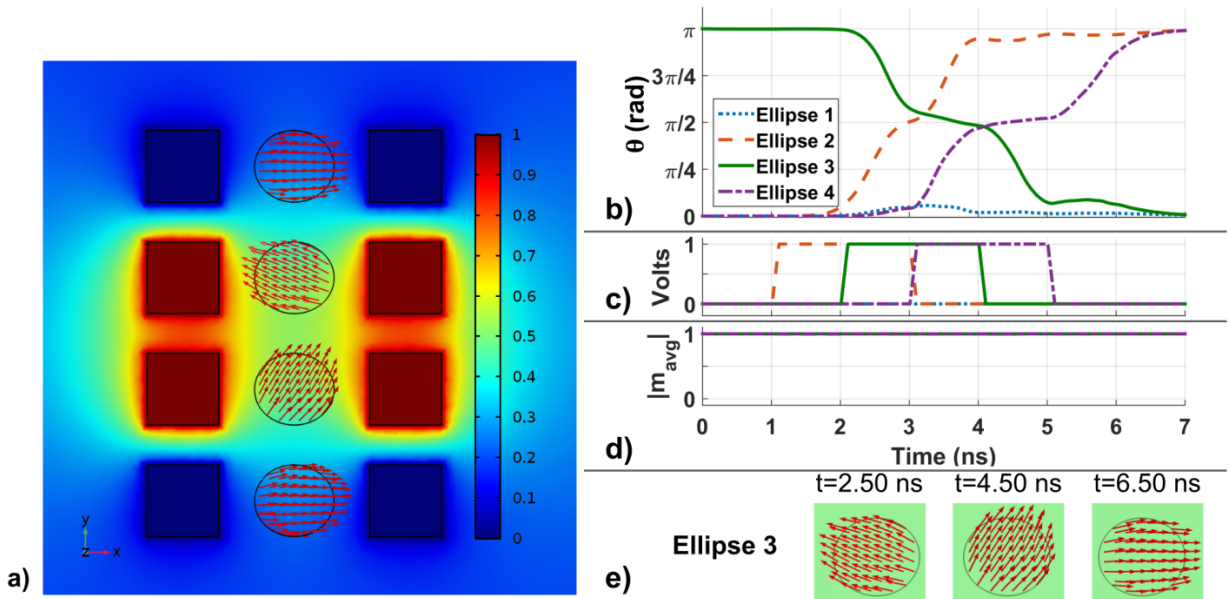

**Figure 4.** Results for Nickel Bennett clocking. (a) Entire Nickel Bennett clocking wire at t=2.5ns. (b) Magnetization rotation angles of four ellipses. (c) Voltage on four electrode pairs. (d) Maximum average magnetization component in four ellipses. (e) Highlight ellipse 3 at t=2.5, 4.5 and 6.5 ns.

Figure 5a shows the entire Terfenol-D Bennett clocking wire at t=2.5 ns. The magnetic orientation is indicated with red arrows, while the surface color shows the electric potential. When voltage is applied to the relevant electrode pairs, a compressive strain is generated in the x-direction, creating a magnetoelastic easy axis in the y-direction. The applied voltage is of the opposite polarity but same magnitude from the Nickel study (i.e., -1V). A key difference between the Nickel and Terfenol-D studies is evident in this figure, with ellipse 3 clearly rotating in an incoherent manner.

Figures 5b and 5c shows a summary of the entire Bennett clocking sequence for the Terfenol-D ellipses. Figure 5b shows the average rotation angle ($\theta$) for each of the four ellipses as a function of time. Figure 5c shows the voltage on each electrode pair as a function of time. Each ellipse shows the same three step switching behavior seen in the Nickel results. While each ellipse has primarily rotated to the new equilibrium configuration by t=7ns, ellipses 2, 3, and 4 are still approximately 20 degrees from the desired final orientation. Ellipse 2 takes approximately 1.5 ns to initially rotate 90 degrees, but only an additional 0.5 ns to reach its final equilibrium position. Ellipse 3 takes 0.5 ns for both stages of rotation, and ellipse 4 initially takes 0.5 ns, followed by nearly 2ns to reach its final equilibrium. It should be noted that the Terfnol-D rotation speed is nearly twice as fast as the fastest Nickel rotation (i.e., 0.5 ns vs. 1.0 ns).

Figures 5d and 5e show the switching in Terfenol-D ellipses was predominantly an incoherent process. The incoherent and vortex phenomenon have been observed in several papers. [30,31] Figure 5d shows that each ellipse becomes non-uniformly magnetized when the strain is applied or released. Non-uniformities lead to incoherent rotation due to the uniaxial nature of magnetoelastic coupling (i.e., up and down are both easy directions when strained). Figure 5d shows that even before the first voltage is applied at 1ns, the average moment of ellipse 2 has clearly fallen below 1. This initial nonuniform magnetization occurs due to larger dipolar coupling between the ellipses and internal demagnetization energy as compared to the one experienced with Nickel. The $M_s$ of Terfenol-D is 1.6 times larger than Nickel's, but the exchange constant is nearly equal. This increases the ratio of demagnetization and dipolar energy to the exchange anisotropy and leads to a nonuniform ground state. Figure 5e shows that with the existence of residual strain caused by magnetoelastic, effects at t=2 ns, ellipse 3 is non-uniformly magnetized, with the left half canted up and right half canted down. Given that $B1 = B2 = \frac{3E\lambda_s}{2(1+\nu)}$, where $B1$ and $B2$ are the magnetoelastic coupling coefficients[29], considering Terfenol-D's extremely high $\lambda_s$, Terfenol-D's magnetization will be very sensitive to any input strain. Because of the non-straight initial state (left half canted up and right half canted down), magnetoelastic anisotropy forces the left half to point up and the right half to point down when the ellipse is subsequently strained, creating an unstable domain wall in the ellipse (t=2.25ns frame of Figure 5e). When this happens, every single magnetization spin finds its shortest way to reach the local minimum state of total energy.[8] After some time, the dipole-dipole effect between magnets forces the magnetization spins to come to the global minimum point of the total energy and switching is completed. The total switching time is then dependent on the time it takes the domain-wall-like magnetization to propagate across the ellipse and restore a stable equilibrium configuration. Propagation of the unstable domain wall explains the initially slow 1.5 ns switching time for ellipse 2. At t=7 ns, the ellipses have rotated the majority of the way into a new antiferromagnetic ground state but are approximately 20 degrees from the desired alignment. Importantly, these observations cannot be made with the macrospin models prevalent in this field, and instead, requires analysis of the spatial and temporal magnetization distribution in each ellipse. This model simulates a similar recent work by Atulasimha et. al[31,32] but also including strain feedback. by modeling nonuniform elastodynamic interactions in each ellipse, instead applying spatially uniform uniaxial anisotropies to simulate strain.

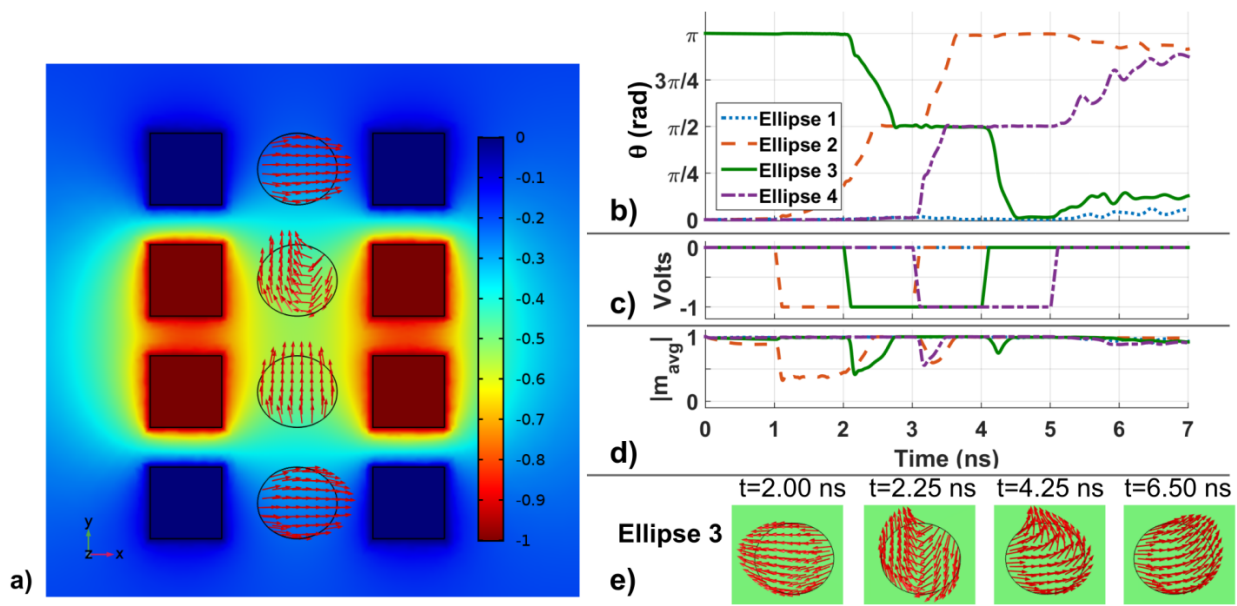

**Figure 5.** Results for Terfenol-D Bennett clocking. (a) Entire Terfenol-D Bennett clocking wire at t=2.5ns. (b) Magnetization rotation angles of four ellipses. (c) Voltage on four electrode pairs. (d) Maximum average magnetization component in four ellipses. (e) Highlight ellipse 3 at t=2.0, 2.25, 4.25 and 6.5 ns.

It should be highlighted that the incoherent rotation seen in the Terfenol-D ellipses can be ameliorated with further device design. By shrinking the overall dimensions of each ellipse and adjusting the inter-ellipse spacing, both the demagnetization and dipolar energies can be reduced to achieve the uniform magnetization profiles seen in the Nickel study. This will reduce the time for unstable domain propagation out of the ellipse and further improve switching times and device reliability.

## 5. Conclusions
This paper uses a numerical model which fully couples elastodynamics and micromagnetics to examine Nickel and Terfenol-D Bennett clocking wires. Nickel ellipses rotated in a coherent manner and produced a stable antiferromagnetically coupled ground state after the wire was clocked. Terfenol-D ellipses were capable of rotating nearly twice as fast as Nickel but were also highly susceptible to incoherent rotations. The larger saturation magnetization for Terfenol-D resulted in non-uniformly magnetized equilibrium states that subsequently led to incoherent rotations. The observations made in this study highlight the need for advanced models that examine the spatial magnetization distribution, in contrast to macrospin modeling. Finally, it was hypothesized that decreasing the size of the Terfenol-D ellipses and altering the inter-ellipse distance will prevent the nonuniformities from initiating incoherent behavior. The incoherent flip indicates that because of the large saturation magnetization of Terfenol-D, the system is much more sensitive to strain compared with Ni system. This should facilitate the use of Terfenol-D in Bennett clocking devices.

**Acknowledgements**

This work was supported by NSF Nanosystems Engineering Research Center for Translational Applications of Nanoscale Multiferroic Systems (TANMS) Cooperative Agreement Award (No. EEC-1160504).